\documentstyle[11pt,IAUS215,twoside,epsf]{article}

\def\edcomment#1{\iffalse\marginpar{\raggedright\sl#1\/}\else\relax\fi}
\reversemarginpar

\begin{document}
\vspace*{1cm}
\title{WIYN Open Cluster Study: Tidal Interactions in Solar type Binaries}
 \author{S. Meibom, R. D. Mathieu}
\affil{University of Wisconsin, Department of Astronomy, Madison WI 53706  USA}

\begin{abstract} We present an ongoing study on tidal interactions
in late-type close binary stars. New results on tidal circularization
are combined with existing data to test and constrain theoretical
predictions of tidal circularization in the pre-main-sequence
(PMS) phase and throughout the main-sequence phase of stellar
evolution. Current data suggest that tidal circularization during
the PMS phase sets the tidal cutoff period for binary populations
younger than $\sim$ 1 Gyr. Binary populations older than $\sim$
1 Gyr show increasing tidal cutoff periods with age, consistent
with active main-sequence tidal circularization.

\end{abstract}

\section{The effects of tidal interactions}

Recent studies of angular momentum gain and loss in young solar-type
stars have primarily been focused on single stars and the interaction
with circumstellar disks (e.g. Barnes et al. 2001, Terndrup et al. 2000,
and Krishnamurthi et al. 1997). Motivated by the high frequency of
late-type binary stars, we will study the affect of close  binary
companions on stellar angular momentum evolution through tidal interactions.

Tidal deformation of stars in binaries results in a torque component
in their gravitational attraction. This torque is responsible for the
spin-orbit coupling by which angular momentum is exchanged between the
star and the orbit (Zahn 1966,1977, Hut 1981 and references therein). Tidal
interactions will continually change the rotation of the binary components
and the orbital parameters of the binary system until an equilibrium
state is reached. An equilibrium state is characterized by co-planarity
of the equatorial planes of the two stars with the orbital plane of the
system, circularity of the stellar orbits, and synchronization of the stellar
rotation with the orbital revolution of the system.

Figure 1 (Zahn \& Bouchet 1989) shows an example of a theoretical
model of tidal evolution from the birth-line through the main-sequence
phase for a binary system of two $1~M_{\sun}$ components. In this model,
the friction that the tidal bulge experiences as the star rotates is caused
by the interaction between convective turbulence and the tidal flow.
The turbulent viscosity is reduced when the orbital period becomes
shorter than the convective turnover time. The torque ($\tau$) on the tidal
bulge is a sensitive function of the ratio of the stellar radius ($R$)
to the distance $r$ between the stars ($\tau \propto \bigl( \frac{R}{r}
\bigr)^{6}$). The timescales for tidal evolution are therefore
extremely sensitive to the stellar radius and the depth of the surface
convection zone. Thus the rate of tidal circularization and synchronization
is higher during pre-main-sequence phase $(t \la 10^{6}~yrs)$ than during
the main-sequence phase. Indeed, Zahn \& Bouchet find that the
main-sequence rate is so small as to not cause any significant tidal
circularization during the main-sequence. The deviation from synchronization
after $10^{6}~yrs$ is due to conservation of angular momentum as the
stars contract before settling on the main-sequence.
 
\begin{figure}[th]
\plotone{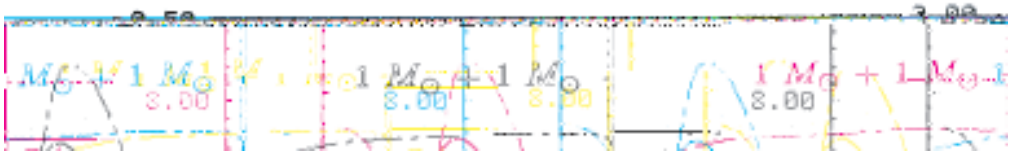}
\caption{Evolution in time of the eccentricity $e$, the orbital
period $P$, and of the ratio between the rotational and orbital
velocities $(\Omega/\omega)$, for a system with two components
of 1 $M_{\odot}$. Note that most of the circularization occurs
well before the ZAMS (indicated by the arrow) while the stars are
still contracting on the Hayashi track. Notice also the synchronism
before the ZAMS and the departure from synchronism as the stars
approach the ZAMS.}
\end{figure}

\section {Observing tidal evolution}

Identification and observations of binary populations of different ages
are needed to test the physics of tidal interactions, and constrain and
direct models of tidal evolution.

Specifically, to study the the rate and evolution of tidal circularization,
observations of orbital eccentricity ($e$) as a function of orbital period
($P$) and stellar age are necessary (see Mathieu et al. 1992).
For a close binary system determination of the orbital parameters can
be obtained through repeated measurements of the radial velocity of one
or both stars. Our radial velocity survey of open clusters with ages from
$\sim$ 3 Myr to $\sim$ 7 Gyr has been a part of the WIYN Open Cluster
Study (WOCS) since 1996. This survey now holds $\sim 20,500$ spectra
of $\sim 4,300$ stars in 6 open clusters. More than 100 binary orbits
have already resulted from this survey, some of which are presented below.

To study the rate and evolution of tidal synchronization, observations
of the rotational periods of stars in binaries must be combined with
determinations of binary orbital parameters for a binary sample of
varying stellar separation (orbital period) and age (see Claret \& Cunha
1997). Stellar rotation periods can be determined from periodic variations
in stellar light-curves caused by star-spots (Barnes et al. 1999, Stassun
et al. 1999). Such photometric time-series studies are underway for the
open clusters M35 (age $\sim$ 0.15 Gyr) and M34 (age $\sim$ 0.25 Gyr)
based on observations with the WIYN 0.9m telescope. A comprehensive
time-series study of the PMS cluster NGC2264 (age $\sim$ 3-5 Myr)
already exists (Lamm et al. 2003, Makidon et al. 2003).

\section {New results on tidal circularization}

Figure 2 shows orbital data for binaries in the two open clusters M35
(left) and NGC188 (right, age $\sim$ 7 Gyr). The plots present orbital
eccentricity as a function of log orbital period for 37 main-sequence
spectroscopic binaries in M35 and 28 main-sequence spectroscopic binaries
in NGC188. In both clusters the shortest period binaries have circular
orbits, while the longest period orbits are all eccentric. The well-defined
transition between circular and eccentric orbits was established by the
CORAVEL (Mayor 1985) and CfA (Mathieu \& Mazeh 1988) teams in the mid
1980's, and led to the definition of the longest period circular orbit
as the {\it "tidal cutoff period"} ($P_{C}$, Duquennoy et al. 1992).
The tidal cutoff period in M35 ($P_{C} = 10.3$ days) and in NGC188
($P_{C} = 14.9$ days) are marked by vertical dashed lines in Fig. 2.
The tidal cutoff period is a measure of the tidal circularization rate as
a function of binary period integrated over the lifetime of the binary
population. 

\begin{figure}[ht]
\plottwo{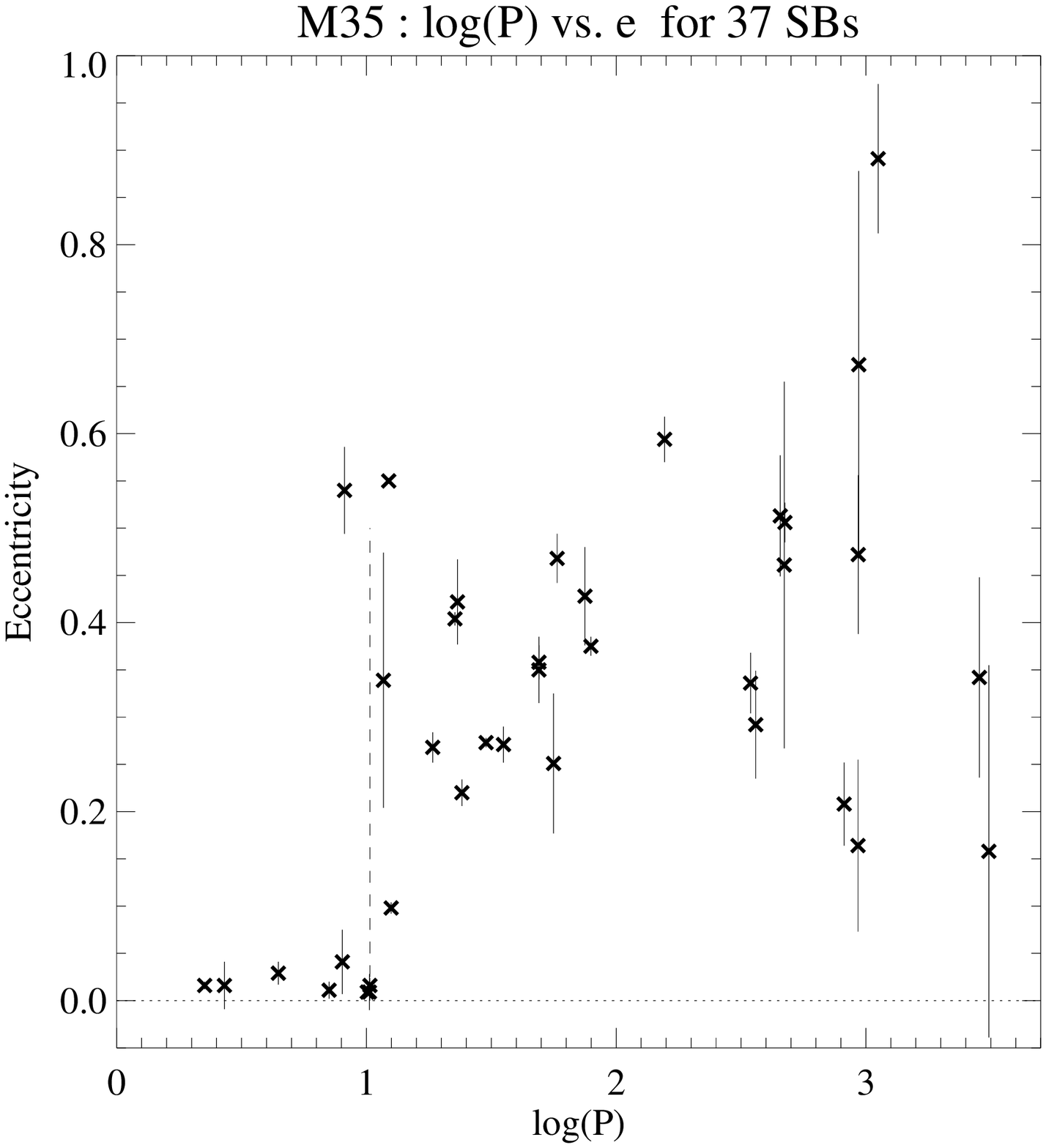}{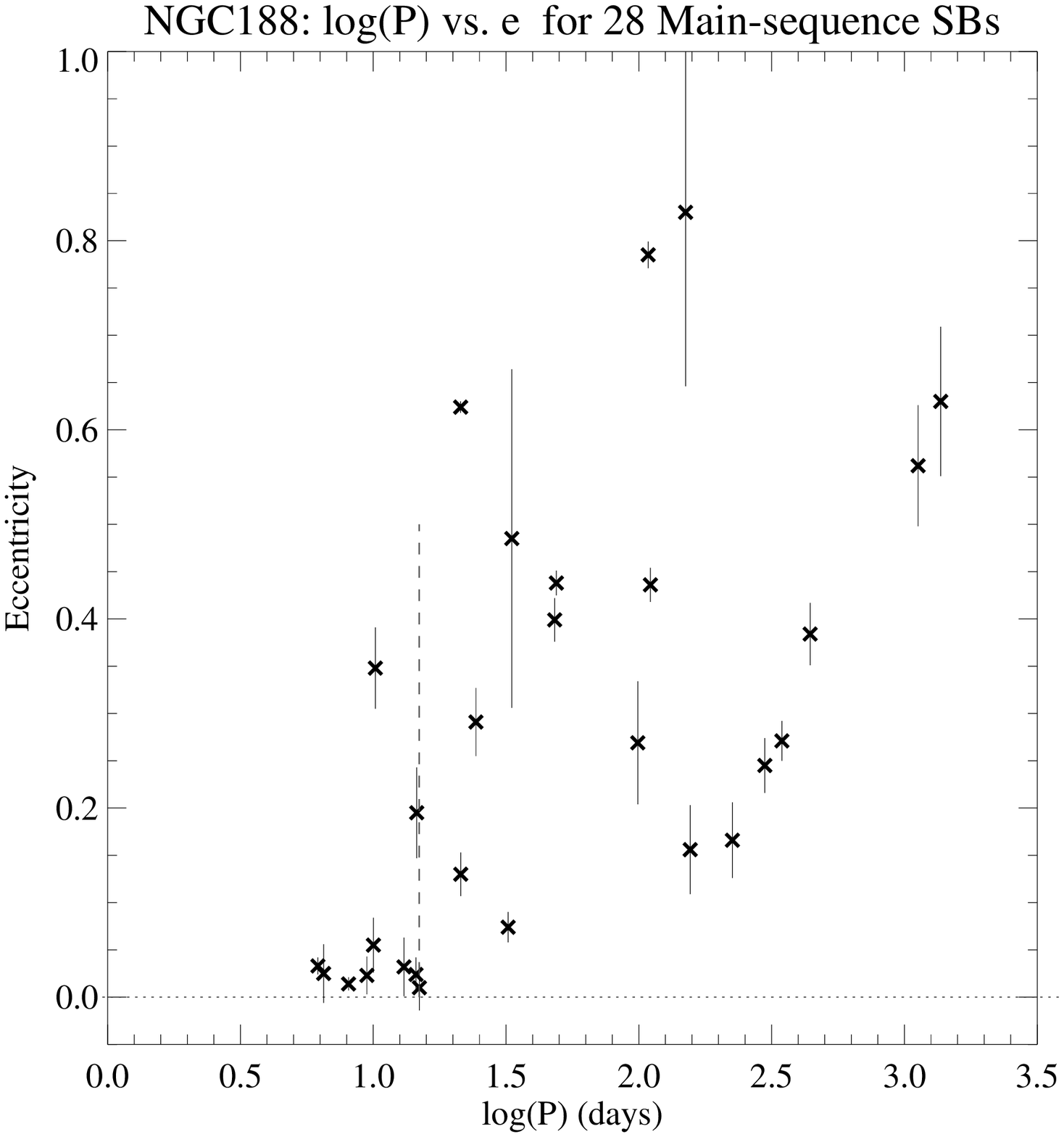}
\caption{Period (log P) vs. eccentricity for 37 main-sequence spectroscopic
binaries in M35 (left) and 28 main-sequence spectroscopic binaries in
NGC188 (right). The vertical dashed lines mark the tidal cutoff periods
of 10.3 days in M35 and 14.9 days in NGC188, respectively}
\end{figure}

\section {Tidal circularization on the main-sequence}

The tidal cutoff periods for several populations of solar-type binaries
has been established over the last two decades. The distribution of
tidal cutoff periods with age enables us to study the evolution of
tidal circularization from the PMS phase (Melo et al. 2001) and through
the main-sequence (Mathieu et al. 1992) and beyond (Verbunt
\& Phinney 1995).  Figure 3 shows our newly derived tidal cutoff
periods for M35 and NGC188 (black circles) combined with previously
determined cutoff periods (grey circles: 7.7 days among PMS binaries (age
$\sim$ 1 Myr, Melo et al. 2001), 7.0 days in the Pleiades (age $\sim$
0.1 Gyr, Duquennoy et al. 1992), 8.5 days in the Hyades/Praesepe (age $\sim$
0.6 Gyr, Duquennoy et al. 1992), 12.4 days in M67 (age $\sim$ 4 Gyr, Latham
et al. 1992a), and 18.7 days for the halo sample (age $\sim$ 13 Gyr, Latham
et al. 1992b).

Concurrent with these observations of binary populations, different
scenarios have been proposed for the evolution of tidal circularization
through the main-sequence:

{\it "No main-sequence tidal circularization"}: This model by Zahn
\& Bouchet (1989) (cf. Fig. 1, horizontal dashed line in Fig. 3) suggests
that PMS tidal circularization sets a tidal cutoff period of $\sim$ 7-8
days and that there is little further circularization throughout the
main-sequence phase. Existing data for PMS binaries support a tidal cutoff
period of $\sim$ 7-8 days. However, data for binary populations older
than $\sim$ 1 Gyr clearly show an increase in the tidal cutoff period
with age, suggesting significant active main-sequence tidal circularization.

{\it "Hybrid scenario"}: While the distribution of tidal cutoff periods
continue to suggest active main-sequence circularization beyond $\sim$
1 Gyr, they do not exclude the idea that PMS circularization sets the
tidal cutoff period for populations younger than $\sim$ 1 Gyr.
The hybrid scenario suggest that the cutoff period of binary populations
younger than $\sim$ 1 Gyr derive from PMS tidal circularization, and that
after the passage of $\sim$ 1 Gyr the integrated main-sequence tidal
circularization begins to circularize binaries with orbital periods
of $\sim$ 7-8 days (Mathieu et al. 1992).

\begin{figure}[!ht]
\plotone{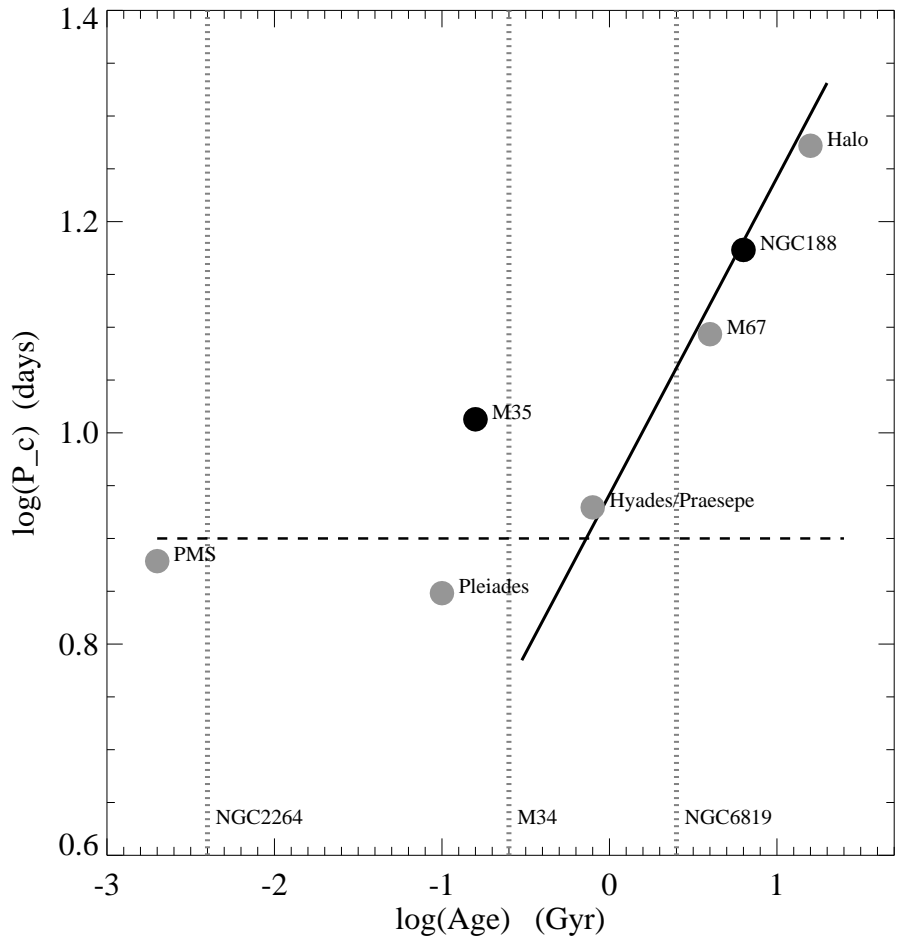}
\caption{Newly derived tidal cutoff periods (black circles) combined
with previously determined cutoff periods (grey circles). The horizontal
dashed line represent the model by Zahn \& Bouchet (1989). The solid
black power-law is a fit to the tidal cutoff periods for binary
populations with ages larger than $\sim$ 1 Gyr, representing a
main-sequence tidal circularization rate proportional to the binary
period to the power of 10/3 (Goldman \& Mazeh 1991). The 3 vertical dashed 
lines indicate the location (in age) of cutoff periods that will be
derived in the near future from the open clusters NGC2264, M34, and NGC6819.
}
\end{figure}

The observed tidal cutoff periods for PMS binaries and binaries in
the Pleiades and Hyades/Praesepe are in agreement with the hybrid scenario,
while the most recent cutoff period in M35 is curiously high by 2-3 days.
The significance of this result is not yet clear. While the tidal cutoff
periods for M35 and the Hyades/Praesepe are well defined, the cutoff
periods for the PMS and Pleiades binary populations are likely lower limits.
The cutoff periods for M35 and Hyades/Praesepe might be showing a scatter
among tidal cutoff periods due to astrophysical differences in the two
binary populations. 

New models for the tidal evolution in eccentric late-type main-sequence
binaries was recently published by Witte \& Savonije (2002) and Savonije
\& Witte (2002). The tidal dissipation in these models is calculated in
the framework of resonant interactions between the harmonic oscillations
created by the perturbing potential of the companion star and the stellar
eigenmode oscillations. PMS tidal evolution was not considered in these
models. At the time of writing this paper we have have not compared
these models to the current set of observed tidal cutoff periods.

Tidal cutoff periods for binary populations at ages: $\sim$ 3-5 Myr
(NGC2264), $\sim$ 250 Myr (M34), and $\sim$ 2.5 Gyr (NGC6819) will be
derived in the near future from ongoing radial velocity surveys on
these clusters. The location (in age) of these future cutoff periods
are indicated by vertical grey dashed lines in Fig. 3.

\section{Tidal synchronization}

The amount of observational data suitable for testing the rate of
synchronization in main-sequence late-type close binaries is sparse.
Claret \& Cunha (1997a,b) analyzed the validity of the theories of
Zahn (1977,1989) and Tassoul \& Tassoul (1990,1992) using the binary
data from Andersen 1991. Most of the work on synchronization
in close binaries has concentrated on early type stars with radiative
envelopes where the proposed mechanism for tidal dissipation is
significantly different (Witte \& Savonije 2001,1999, Giuricin et al.
1984a,b).

Our immediate goal is to study synchronization during the PMS and
early main-sequence (age $\la$ 300 Myr) phases by comparing the
rotational periods of stars in binary systems with their orbital
velocity.

\section {Conclusions}

Binary populations younger than $\sim$ 1 Gyr have tidal circularization
cutoff periods between $\sim$ 7-10 days suggesting that a 7-10 day tidal
cutoff period is set during PMS phase. After the passage of $\sim$ 1 Gyr
tidal cutoff periods increase with increasing age suggesting active
main-sequence circularization. This distribution of tidal cutoff periods
with age is in agreement with the hybrid scenario proposed by Mathieu et al.
(1992). However, there are currently no theoretical model for tidal
circularization that can account for both the observed PMS and main-sequence
evolution of tidal cutoff periods.

\end{document}